\documentclass[prd,reprint,aps,tightenlines,preprintnumbers,amsmath,amssymb,
showpacs,superscriptaddress]{revtex4-1}
\usepackage{amsmath,amssymb,graphicx,natbib}
\usepackage{graphicx}
\usepackage{dcolumn}
\usepackage{bm}
\usepackage[hyperindex,breaklinks]{hyperref}
\usepackage{amsmath}
\usepackage{array}

\def\be{\begin{equation}}
\def\ee{\end{equation}}
\def\ba{\begin{eqnarray}}
\def\ea{\end{eqnarray}}
\def\ge{\mathrel{\raise.3ex\hbox{$>$\kern-.75em\lower1ex\hbox{$\sim$}}}}
\def\la{\mathrel{\raise.3ex\hbox{$<$\kern-.75em\lower1ex\hbox{$\sim$}}}}

\def\simgt{\mathrel{\raise.3ex\hbox{$>$\kern-.75em\lower1ex\hbox{$\sim$}}}}
\def\simlt{\mathrel{\raise.3ex\hbox{$<$\kern-.75em\lower1ex\hbox{$\sim$}}}}

\newcommand{\nc}{\newcommand}

\nc{\gone}{\bar g_{\pi NN}^{(1)}}
\nc{\gzero}{\bar g_{\pi NN}^{(0)}}
\nc{\al}{\alpha}
\nc{\ga}{\gamma}
\nc{\de}{\delta}
\nc{\ep}{\epsilon}
\nc{\ze}{\zeta}
\nc{\et}{\eta}

\nc{\Th}{\Theta}
\nc{\ka}{\kappa}
\nc{\rh}{\rho}
\nc{\si}{\sigma}
\nc{\ta}{\tau}
\nc{\up}{\upsilon}
\nc{\ph}{\phi}
\nc{\ch}{\chi}
\nc{\ps}{\psi}
\nc{\om}{\omega}
\nc{\Ga}{\Gamma}
\nc{\De}{\Delta}
\nc{\La}{\Lambda}
\nc{\Si}{\Sigma}
\nc{\Up}{\Upsilon}
\nc{\Ph}{\Phi}
\nc{\Ps}{\Psi}
\nc{\Om}{\Omega}
\nc{\ptl}{\partial}
\nc{\del}{\nabla}
\nc{\ov}{\overline}
\nc{\newcaption}[1]{\centerline{\parbox{15cm}{\caption{#1}}}}

\def\beq{\begin{equation}}
\def\eeq{\end{equation}}
\def\bmat{\begin{displaymath}}
\def\emat{\end{displaymath}}
\def\bear{\begin{eqnarray}}
\def\eear{\end{eqnarray}}
\def\bery{\begin{array}}
\def\ery{\end{array}}
\def\bit{\begin{itemize}}
\def\eit{\end{itemize}}
\def\ben{\begin{enumerate}}
\def\een{\end{enumerate}}
\def\btab{\begin{tabular}}
\def\etab{\end{tabular}}
\def\btbl{\begin{table}}
\def\etbl{\end{table}}
\def\bfig{\begin{figure}[htb]}
\def\efig{\end{figure}}
\def\bpic{\begin{picture}}
\def\epic{\end{picture}}

\def\ga{\mathrel{\raise.3ex\hbox{$>$\kern-.75em\lower1ex\hbox{$\sim$}}}}
\def\la{\mathrel{\raise.3ex\hbox{$<$\kern-.75em\lower1ex\hbox{$\sim$}}}}
\def\gappeq{\mathrel{\rlap {\raise.5ex\hbox{$>$}}
{\lower.5ex\hbox{$\sim$}}}}
\def\lappeq{\mathrel{\rlap{\raise.5ex\hbox{$<$}}
{\lower.5ex\hbox{$\sim$}}}}

\def\gyr{{\rm \, G\kern-0.125em yr}}
\def\mev{{\rm \, Me\kern-0.125em V}}
\def\gev{{\rm \, Ge\kern-0.125em V}}
\def\tev{{\rm \, Te\kern-0.125em V}}

\begin{document}

\title{Constraints on muon-specific dark forces}

\author{Savely~G.~Karshenboim}
\affiliation{Max-Planck-Institut f\"ur Quantenoptik, Garching, 85748, Germany}
\affiliation{Pulkovo Observatory, St.~Petersburg, 196140, Russia}

\author{David~McKeen}
\affiliation{Department of Physics, University of Washington, Seattle, WA 98195, USA}

\author{Maxim~Pospelov}
\affiliation{Department of Physics and Astronomy, University of Victoria, 
Victoria, BC V8P 5C2, Canada}
\affiliation{Perimeter Institute for Theoretical Physics, Waterloo, ON N2J 2W9, 
Canada}

\begin{abstract}

The recent measurement of the Lamb shift in muonic hydrogen allows for the most precise extraction of the 
charge radius of the proton which is currently in conflict with other determinations based on
$e-p$ scattering and hydrogen spectroscopy. This discrepancy could be the result of some
new  muon-specific force with O(1-100) MeV force carrier---in this paper we concentrate on vector mediators. 
Such an explanation faces challenges from the 
constraints imposed by the $g-2$ of the muon and electron as well as precision spectroscopy of  muonic atoms. 
In this work we complement the family of constraints by calculating the contribution of 
hypothetical forces to the muonium hyperfine structure. We also compute the two-loop contribution to the 
electron parity violating amplitude due to a muon loop, which is sensitive to the muon axial-vector coupling. 
Overall, we find that the combination of low-energy constraints favors the mass of the mediator to be below 10 MeV, and 
that a certain degree of tuning 
is required between vector and axial-vector couplings of 
new vector particles to muons in order to satisfy constraints from 
muon $g-2$. However, we also observe that in the absence of a consistent standard model embedding, high energy weak-charged 
processes accompanied by the emission of new vector particles are strongly enhanced by $(E/m_V)^2$, 
with $E$ a characteristic energy scale and $m_V$ the mass of the mediator. In particular, leptonic $W$  decays impose the strongest 
constraints on such models completely disfavoring the remainder of the parameter space. 


\end{abstract}

\maketitle

\section{Introduction}

The persistent discrepancy of the measured muon $g-2$ and the standard model (SM) prediction at the 
level of $\sim$3$\sigma$~\cite{Bennett:2006fi} has generated a lot of experimental and theoretical activity in search of a possible 
explanation. Among the new physics explanations for this discrepancy are weak scale solutions~\cite{Czarnecki:2001pv} 
{\em and} possible new contributions from light and very weakly coupled new particles (see, e.g.,~\cite{Pospelov:2008zw}). For the latter case 
there must be additional observable effects that involve muons and new forces mediated by light particles. 

Recently, a new intriguing discrepancy has emerged after the Lamb shift in muonic hydrogen 
has been measured at PSI. 
The 2010-2012 results~\cite{Pohl:2010zza,Antognini:1900ns} 
combined with the QED calculations of the same quantity allow for a very accurate extraction
of the charge radius of the proton. The result stands in sharp contradiction with the 
determination of the proton charge radius in electron-proton scattering experiments and from 
high-precision spectroscopy of ``normal" hydrogen and deuterium, as summarized in the CODATA 
review~\cite{Mohr:2012tt}. 
The combined discrepancy stands now at more than $7\sigma$, with $5\sigma$ discrepancies with H spectroscopy and scattering separately, and therefore should be taken very seriously. 
Unlike the case of the $g-2$ discrepancy, this latest contradiction {\em cannot} be a result of new physics at the weak scale. 

Broadly speaking, 
there are several logical pathways toward resolving the present contradiction:
\begin{enumerate}

\item    The muonic atom results are obtained by only one group, and could contain an unaccounted source of error. 
However, so far no credible candidates for a systematic shift on the order  of $0.3$ meV have been found. Moreover, the measurement 
of two lines in muonic hydrogen exhibit full self-consistency~\cite{Pohl:2010zza,Antognini:1900ns}. 
At the level of accuracy set by the current size of the discrepancy, $\delta E \sim 0.3~{\rm meV}$, the 
QED part of the muonic hydrogen Lamb shift calculation is comparatively simple and 
has been checked by many groups. For a compilation of the related theoretical issues see Ref.~\cite{Antognini:2012ofa}.

\item   Strong interactions could affect the Lamb shift in $\mu$H via a two-photon polarization diagram.
Standard calculations based on a dispersive approach (see {\em e.g.}~\cite{Carlson:2011zd} for the latest evaluations) 
show no room for a contribution that could account for the discrepancy. 
Still, some of the input to these calculations has model dependence built-in~\cite{Hill:2011wy}, and exaggerating this dependence 
to the extreme~\cite{Miller:2012ne} could hypothetically provide a large frequency shift. In this case, however, one should expect
drastic deviations for the hadronic two-photon effects elsewhere~\cite{WalkerLoud:2012bg} which are not observed. 
Therefore, this is also an unlikely 
proposition. 

\item  The problem could lie with the determination of $r_p$ in standard hydrogen. Notice that in order to be 
consistent with the muonic hydrogen Lamb shift, results based on {\em both} methods, $e-p$ scattering and hydrogen spectroscopy, 
would have to be incorrect or have overstated precision. 

\item Finally, it is also possible that some ``intermediate range" force is responsible for the discrepancy. 
Should such a new force carrier exist in the MeV-100 MeV mass range, it could potentially 
affect the $\mu$H Lamb shift directly. Constructing a model that would be not immediately ruled out by the existing 
constraints on dark forces in this range is a difficult challenge~\cite{Barger:2010aj,TuckerSmith:2010ra,Batell:2011qq}. 

\end{enumerate}

Further background information and discussion can be found in the recent review~\cite{Pohl:2013yb}.

\begin{figure}
\begin{center}
\resizebox{0.9\columnwidth}{!}{\includegraphics{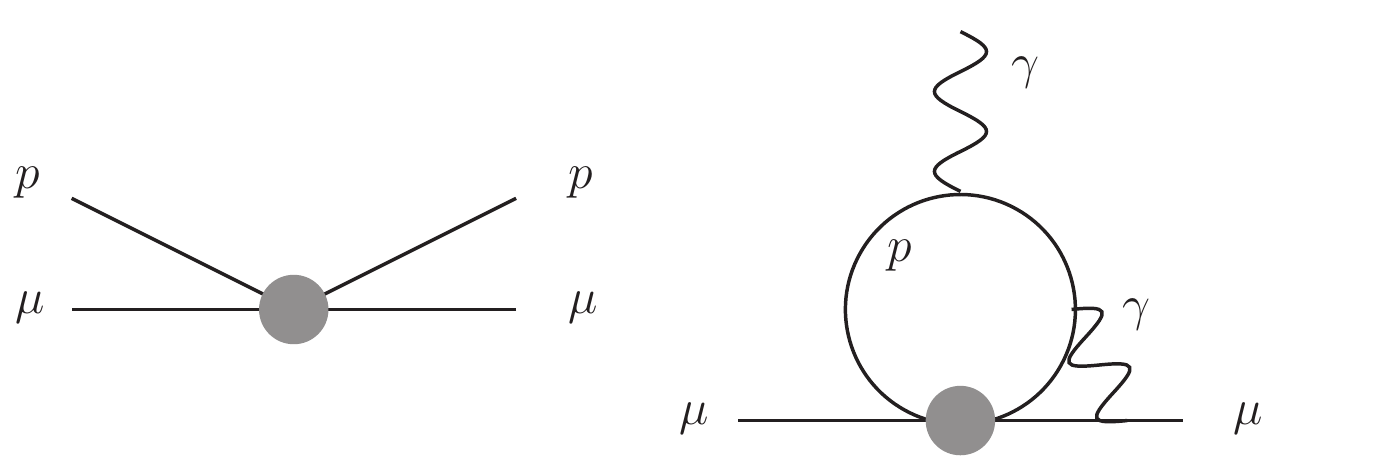}}
\caption{Left:the effective proton-muon interaction resulting from unexpectedly large QCD effects or new physics that is responsible for the $r_p$ discrepancy. Right: the two-loop contribution to the muon $g-2$ that results from the interaction on the left after integrating out the proton.}
\label{f:naive}
\end{center}
\end{figure}

The search for a resolution to the $r_p$ discrepancy is important because it carries strong implications for the precision of
 theoretical evaluation of the muon 
$g-2$. Suppose, for example, that either ``unexpected" effects of strong interactions (solution 2 above), or 
some new physics (solution 4) is responsible for inducing, e.g., a large proton-muon interaction term,
\be
\Delta {\cal L} \simeq C(\bar \psi_\mu \psi_\mu)(\bar \psi_p \psi_p),
\label{verynaive}
\ee
where coefficient the $C$ needs to be $\sim (4 \pi \alpha)\times 0.01~{\rm fm}^2$ in order to explain the discrepancy
in $r_p$ measurements. This effective interaction is shown on the left of Fig.~\ref{f:naive}. One can then estimate the typical shift to the muon $g-2$ that 
this interaction would imply by integrating out the proton, leading to the two-loop effect on the right of Fig.~\ref{f:naive}. 
(Other charged hadrons presumably would contribute as well.) 
Using (\ref{verynaive}) as a starting point, we perform 
 a simple estimate by rescaling the well-known perturbative formula for the 
two-loop Higgs/heavy quark contributions to the muon $g-2$ found in, e.g.,~\cite{Gnendiger:2013pva}. 
Since we are converting a dimension-6 operator in (\ref{verynaive}) into the dimension-5 $g-2$ operator, the result is 
linearly divergent and presumably is stabilized by some hadronic scale $\Lambda_{\rm had}$, where 
neither the coefficient $C$ nor the proton-photon vertex can be considered local. Taking a wide range for 
$\Lambda_{\rm had}$, from a proton mass scale $m_p$ to a very light dynamical scale $\sim m_\pi$, 
one arrives at the following estimates of a typical expected shift for the muon anomalous magnetic moment, 
\begin{eqnarray} 
\Delta (a_\mu) \sim - C \times \frac{\alpha m_\mu m_p}{8 \pi^3}
\times \left\{\begin{array} {c} 1.7 ;~~\Lambda_{\rm had} \sim m_p\\ 0.08;~\Lambda_{\rm had} \sim m_\pi
  \end{array}\right.,
\end{eqnarray}
which, after inputing the value of $C$ implied by the $r_p$ discrepancy results in
\be
5\times 10^{-9} \la |\Delta (a_\mu) | \la 10^{-7}.
\ee
Clearly, the upper range of this possible shift is enormous while the lower range is 
still large, on the order of the existing discrepancy in muon $g-2$. It is three times the size of the current 
estimates for the hadronic light-by-light contributions, and one order of magnitude larger than the 
uncertainty claimed for that contribution. 
These estimates show that if indeed large muon-proton interactions are responsible for
the $r_p$ discrepancy, one can no longer insist that theoretical 
calculations of the muon $g-2$ are under control. Thus, a resolution 
of the $r_p$ problem is urgently needed in light 
of the new  significant investments made in 
the continuation of  the experimental $g-2$ program. 

In this paper, we entertain the possibility (solution 4) that a new vector force is responsible for the discrepancy. 
Our goal is to investigate the status of this vector force in light of the  $g-2$ results for the electron and muon and to
derive additional  constraints from the hyperfine structure of muonium. As we will show, the presence of a 
parity-violating coupling to the muon is a very likely consequence of such models, and in light of that we calculate the two-loop 
constraint on the parity violating muon-nucleon forces imposed by ultra-precise tests of parity in the electron sector. 
We believe that our analysis is timely, given the new experimental information that will soon emerge from the measurement of the Lamb shift in muonic deuterium and helium and the new efforts at 
making the ordinary hydrogen measurements more precise. 
 
Our approach to the new force is purely phenomenological. At the same time it is important to realize 
that the embedding of such new force into the structure of the SM is very difficult 
and so far no fully consistent models of such new interaction have been proposed. (The closest attempt, 
the gauged $\mu_R$ model of Ref.~\cite{Batell:2011qq}, suffers from a gauge anomaly and thus must be regarded as an 
effective model up to some ultraviolet scale, close to the weak scale.)  Therefore, even a phenomenologically 
successful model that would explain the $r_p$ discrepancy and pass through all additional constraints should be 
viewed at this point as an exercise which can be taken more seriously only if a credible SM embedding is found, or if
the new force hypothesis finds further experimental support.

We illustrate the need for the consistent SM embedding explicitly, by considering the 
high-energy constraints on the muon-specific vector force. We show that normally not-so-precise 
observables such as $W$-boson decay branching fractions become extremely constraining, 
since they are affected by the muon-specific force because of the 
breaking of the full SM gauge invariance. We observe that $\sim (E/m_\mu)^2$ enhancement of all charged current 
effects is a generic price for the absence of a consistent SM embedding, which strongly disfavors such models.

This paper is organized as follows. In the next section we introduce a model for an intermediate-range force, and determine 
the parameter values suggested by the $r_p$ anomaly. In Sec.~\ref{sec:hfs_np} we calculate the one loop contribution to the muonium hyperfine structure.
Section~\ref{sec:2loop} contains the calculation of the two-loop transfer of the parity violation in the muon sector to electrons. 
Section~\ref{sec:high-energy} has a the discussion of the high-energy constraints. 
Section~\ref{sec:combination} combines all the constraints on the model and we reach our conclusions in Sec.~\ref{sec:conclusions}. 

\section{Intermediate-range force}
\label{sec:int_force}

We will choose an entirely phenomenological approach and allow for one new particle to mediate the new force between muons and protons. 
Motivated by dark photon models~\cite{Holdom:1985ag}, we assume that the new particle mostly interacts with the 
electromagnetic current and, in addition, has further vector and axial-vector coupling to muons. 
The interaction Lagrangian for this choice is given by 
\begin{eqnarray}
\nonumber
{\cal L}_{\rm int} = - V_\nu\left[\kappa  J^{\rm em} _\nu - \bar\psi_\mu(g_V \gamma_\nu + g_A \gamma_\nu\gamma_5) \psi_\mu \right]\\
\label{start}
= - V_\nu\left[  e\kappa \bar \psi_p \gamma_\nu \psi_p - e\kappa \bar \psi_e \gamma_\nu \psi_e \right. \\\nonumber \left. -  
\bar\psi_\mu((e\kappa +g_V) \gamma_\nu + g_A \gamma_\nu\gamma_5) \psi_\mu +... \right] ,
\end{eqnarray}
where the last two lines describe interaction of the vector, $V$, with the relevant fields: electron, muon, and proton. 
We use positive $e=(4\pi\alpha)^{1/2}$. 
The constant $\kappa$ is the mixing angle between the photon and $V$. It is a safe assumption that this mixing must be small. 
$g_V$ and $g_A$ are the new phenomenological muon-specific couplings that are introduced in this paper by hand.

The interaction via a conserved current, $\kappa  J^{\rm em} _\nu$ allows for a UV completion via kinetic mixing, and
is totally innocuous. The muon-specific couplings $g_V$ and $g_A$ are much more problematic from the point of view of
UV completion and full SM gauge invariance. Notice that in parallel to the kinetic mixing type coupling $V_\nu\kappa  J^{\rm em} _\nu$,
there exists another ``safe" coupling via the baryonic current, $V_\nu (\bar \psi_p \gamma_\nu \psi_p + \bar \psi_n \gamma_\nu \psi_n)$.
The reason we suppress it in this paper is because of the extra phenomenological problems it creates, chiefly the 
additional O(10-100 fm) range force for neutrons--a possibility that is very constrained by  neutron scattering experiments. 
It may look strange that the new force introduced in (\ref{start}) includes parity violation for muons. In fact, as we will see shortly, the 
$g_A$ coupling is necessary to cancel the excessive one-loop contribution to the muon $g-2$ generated by the $g_V$ coupling.

Having formulated our starting point with the Lagrangian in Eq.~(\ref{start}), it is easy  to present a combination of couplings that alleviates the 
current $r_p$ discrepancy. Choosing the 
same sign for $\kappa$ and $g_V/e$ will create an additional attractive force between protons and muons.
It will be interpreted as the difference between charge radii observed in regular and muonic hydrogen:
\begin{eqnarray}
 \left. \Delta r^2\right|_{\mu \rm H} - \left. \Delta r^2\right|_{\rm H} = -\frac{ 6\kappa(\kappa + g_V/e)}{m_V^2}+\frac{ 6\kappa^2}{m_V^2} \nonumber\\
\label{rp}
=- \frac{ 6\kappa(g_V/e)}{m_V^2}~~~~~~~~~~~ \\
\simeq  -0.06~{\rm fm}^2\times \frac{(20~{\rm MeV})^2}{m_V^2} \times \frac{\kappa}{(3\times 10^{-6})^{1/2}} \times \frac{g_V/e}{0.06}
\nonumber
\end{eqnarray}
Here we explicitly assume that the momentum transfer in the $\mu$H system, $\alpha m_\mu$, is smaller than the 
mass of the mediator, $m_V$.  In the second line we have normalized the coupling in such a way as to factor out the size of 
the suggested correction for $r_p$, which corresponds to a relative shift of the squared radius of $0.06~{\rm fm}^2$. At the same time, we have normalized $m_V$ and $\kappa$ on their values 
that correspond to the borderline of the constraint that comes from combining the electron $g-2$ measurement with QED theory and the independent atomic physics determination of $\alpha$. 

Equation~(\ref{rp}) makes clear the fact that given the strong constraints on $\kappa$ and $m_V$, only relatively large values 
for the muon-specific coupling $g_V$ are capable of correcting the $r_p$ anomaly. At the same time, it is clear that the 
muon $g-2$ value will be in conflict with $g_V \sim 0.06$ {\em unless} there is a significant degree of cancellation between 
$g_V^2$- and $g_A^2$-proportional contributions. Fortunately, such contributions are of the opposite sign and the possibility of cancellation does exist. Moreover, since in the limit of $m_V\ll m_\mu$ the contribution of the axial-vector coupling to anomalous magnetic moment $a_\mu$ is parametrically enhanced 
compared to the vector coupling, 
\begin{eqnarray} 
\frac{\Delta a_{\mu}(g_A)}{\Delta a_{\mu}(g_V)} \simeq - \frac{2 g_A^2}{g_V^2} \times \frac{m_\mu^2}{m_V^2},
\nonumber \\
\Longrightarrow g_A^{\rm tuned}  = \pm g_V \times \frac{m_V}{\sqrt{2} m_\mu},
\label{cancel}
\end{eqnarray}
such a cancellation can be achieved with a relatively small value of $g_A\sim {\rm few} \times 10^{-4}$. 
Such small values of $g_A$ still induce a parity violating amplitude for muons well above the 
level suggested by the weak interactions. However, the direct tests of neutral current parity violation for muons at low energy have not been carried 
out directly~\cite{Missimer:1984hx}, and the existence of enhanced parity-violating effects involving 
muons should be regarded as an opportunity to test these models 
in the future~\cite {McKeen:2012sh}. We also note that the similar tuning of vector against axial-vector contribution is {\em not} possible 
for the electron $g-2$, mainly because of the lack of corresponding enhancement for the axial-vector contribution and excessively strong constraints on 
new axial couplings for electrons. 

Finally, we comment on the possibility that a scalar particle mediates a long-range force. 
On one hand, the constraints from $g-2$ of the electron 
are milder because it is reasonable to expect that the coupling would scale proportional to mass, 
$g_S^e/g_S^\mu \sim m_e/m_\mu$. On the other hand, the coupling to neutrons that would also be a generic consequence of 
such model would limit $g_S^{n,p}$ to below the $10^{-3}-10^{-4}$ level, requiring the coupling to muons be $\sim 10^{-2}$ and larger. 
As in the vector case, the correction to $g-2$ of the muon is too large. Unlike the vector case, one cannot use the opposite 
parity coupling to cancel this contribution. This is because the cancellation can be achieved only when the pseudoscalar 
coupling is approximately the same as the scalar one,  $g_P^\mu \simeq g_S^\mu$. This maximally $CP$-violating case leads to unacceptably large EDMs of neutrons and heavy atoms, even after making generous allowance for the suppression coming from the two-loop 
mediation mechanism. Therefore, one needs {\em extra} light states beyond a single scalar. We therefore abandon this possibility, and concentrate on the vector force (\ref{start}), 
where only one new particle is introduced.

\section{Muonium hfs and new physics}
\label{sec:hfs_np}

The best experimental result on the muonium 1s hyperfine structure (HFS) interval is~\cite{mu1shfs}
\begin{equation}
\nu (1s, {\rm hfs})=4463\,302.776(51)\;{\rm
kHz}\;.\label{exp}
\end{equation}

To compare it with theory one has to find the leading term, the so-called Fermi energy,
\begin{eqnarray}
\frac{E_{\rm F}}{h}&=&\frac{16\alpha^2}{3\pi}\,
\frac{\mu_\mu}{\mu_{\rm B}} \,cR_\infty\left(1+\frac{m_e}{m_\mu}\right)^{-3}\nonumber\\&=& \frac{16\alpha^2}{3\pi}\,
\frac{(1+a_\mu)m_e}{m_\mu}  \,cR_\infty\left(1+\frac{m_e}{m_\mu}\right)^{-3}\;,\label{ef}
\end{eqnarray}
and QED corrections to it~\cite{my_rep,book}. The Fermi energy can be presented in terms of fundamental constants in many different ways, but unavoidably, when describing the HFS interaction of a muon (and electron) one has to input either the 
muon magnetic moment or the muon mass in appropriate units.

Presently, it is the determination of the muon mass (or muon magnetic moment) 
\cite{mu1shfs} that dominates the uncertainty of the theoretical prediction~\cite{my_rep,Mohr:2012tt},
\begin{equation}
\nu (1s, {\rm hfs})=4463\,302.89(27)\;{\rm
kHz}\;,\label{th}
\end{equation}
leading to the following comparison of theory and experiment:
\begin{equation}
\frac{\nu^{\rm exp} -\nu^{\rm th}}{\nu^{\rm exp}} =(-2.5\pm1.2\pm6.1)\times10^{-8}\;.\label{comp}
\end{equation}
The concordance determines a room for possible exotic corrections, which we limit at $2\sigma$, 
\begin{equation}
\label{hfslimit}
\left|  \frac{\Delta E_{\rm hfs} }{E_{\rm hfs}}\right| < 1.24 \times 10^{-7}.
\end{equation}

\begin{figure}
\begin{center}
\resizebox{0.65\columnwidth}{!}{\includegraphics{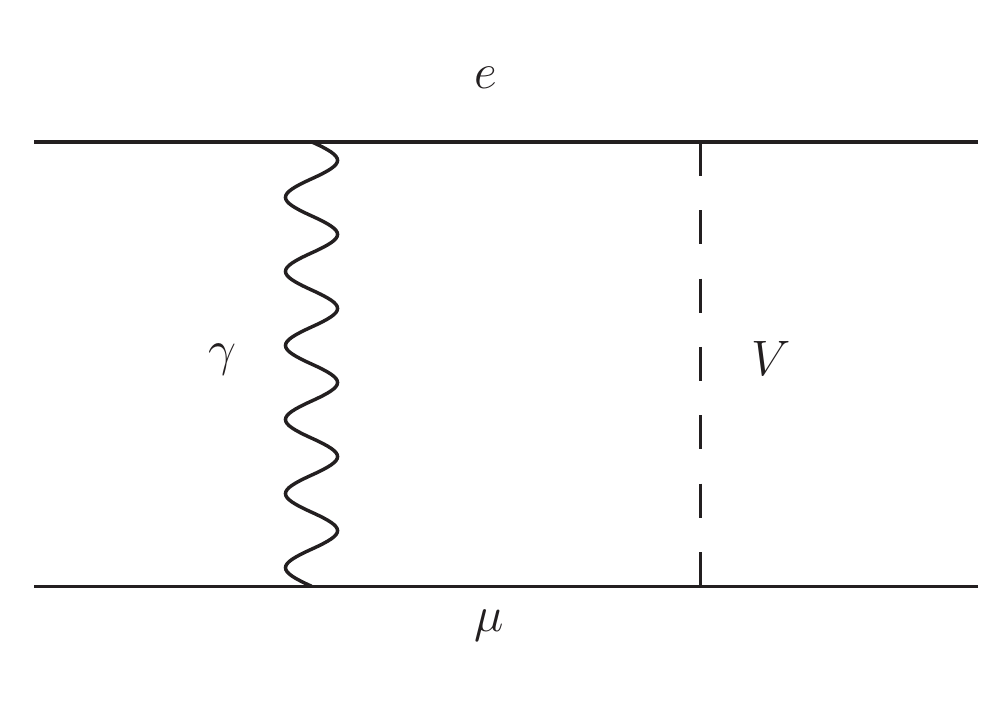}}
\caption{One-loop diagram (plus all possible crossings) contributing at $1/m_V$ order to the muonium HFS.}
\label{f:hfs}
\end{center}
\end{figure}

One has to remember that calculation of the Fermi energy involves fundamental constants and any effect of new physics would affect their determination as well (see, e.g.,~\cite{prl}). Here, we are most interested in the mediator mass range that is 
higher than the typical momenta of atomic constituents, and much higher than that of macroscopic physics.
Therefore, the atomic determination of $\alpha$ and $m_e/m_\mu$ are unaffected by new physics. Indeed, the value for
$m_e/m_\mu$ comes from measurements of the hyperfine structure of muonium in a magnetic field. The magnetic field dependence 
and the determination of the field through free proton precession produce a value for $m_e/m_\mu$, 
which corresponds to very low momentum transfers and 
is not sensitive to short-range effects. Therefore, we can safely proceed by calculating the contributions from the box diagram in Fig.~\ref{f:hfs}. 

In the limit $m_\mu \gg m_V$, the calculation is simple, and we adjust the known formula for the  Zemach correction~\cite{Zemach} in the hydrogen atom to calculate the contribution of the $V$-mediated force in muonium,
\begin{eqnarray}
\frac{\Delta E_{\rm hfs}}{E_{\rm hfs}}\simeq
\frac{2\alpha m_{e\mu}}{\pi^2}\int \frac{d^3p}{p^4}\left[\frac{G_E(-p^2)G_M(-p^2)}{\mu_\mu}-1\right]\!\!,~~~
\end{eqnarray} 
where $p^2$ is the square of the space-like momentum, $m_{e\mu}$ is the reduced mass of the muon and electron, and 
$G_{E(M)}$ are the electric and magnetic form factors. The $V$-mediated Yukawa contribution can be interpreted as effective $G_{E(M)}$ form factors given by
\begin{equation}
G_{E(M)} -1 = \frac{\alpha'}{\alpha} \times \frac{1}{p^2+m_V^2} = \frac{\kappa(\kappa +g_V/e)}{p^2+m_V^2},
\end{equation} 
which defines the exotic coupling $\alpha'$ in our model. Performing the remaining integral, and taking $m_{e\mu} = m_e$, 
one arrives at a rather simple result,
\begin{equation}
\label{resulthfs}
\frac{\Delta E_{\rm hfs}}{E_{\rm hfs}} = \frac{8 \alpha' m_e}{m_V} = \frac{8 \alpha \kappa(\kappa +g_V/e) m_e}{m_V}.
\end{equation}
The full result, without taking $m_V\ll m_\mu$, is derived in the Appendix which exploits the existing 
more precise calculations of the two-photon contribution due to hadronic vacuum polarization~\cite{plb}.
Either way, comparing Eq.~(\ref{comp}) with $r_p$-suggested choice of couplings, Eq.~(\ref{rp}), one can see that the 
muonium HFS provides a nontrivial constraint on the model. 

\section{Two-loop induced PNC amplitude}
\label{sec:2loop}

\begin{figure}
\begin{center}
\resizebox{0.75\columnwidth}{!}{\includegraphics{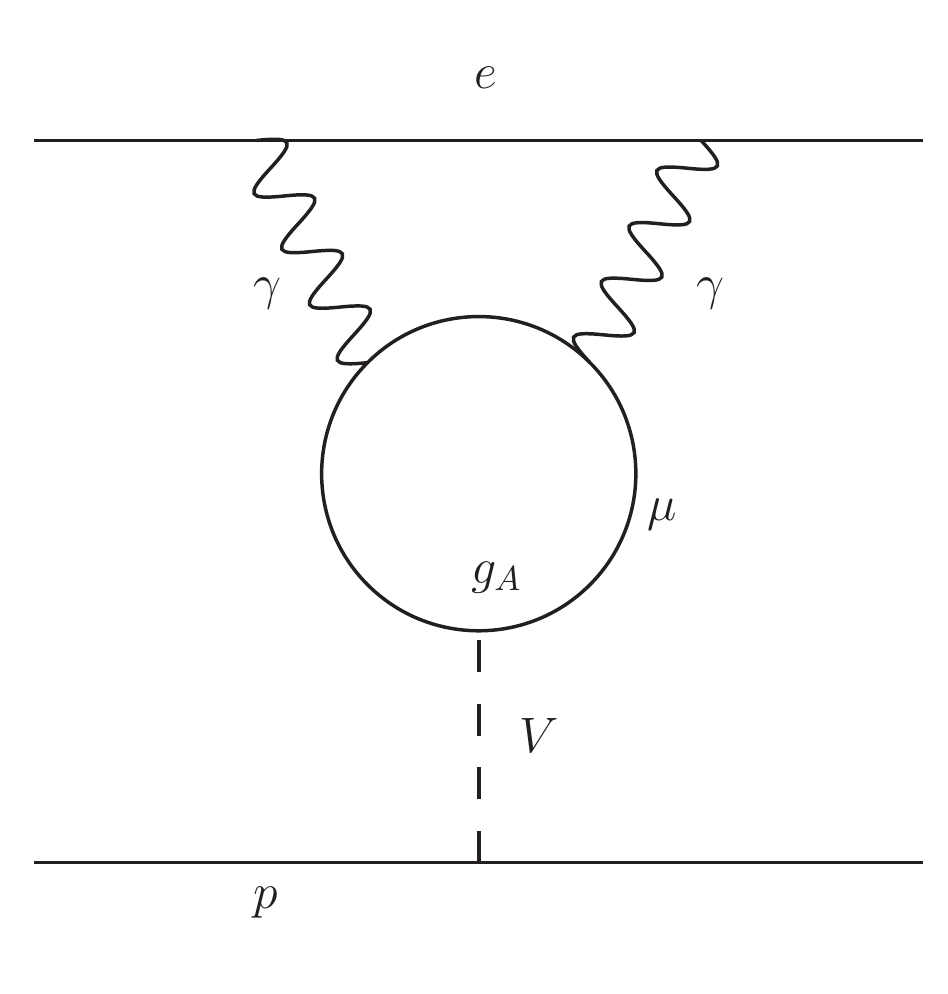}}
\caption{Two-loop diagram with the closed muon loop contributing 
to the atomic PNC amplitude. This diagram does not decouple in the large $m_\mu$ limit.}
\label{fig:PNC}
\end{center}
\end{figure}

The necessity of introducing an axial-vector coupling results in stronger-than-weak amplitudes for 
parity non-conservation (PNC) effects involving muons. However, since there are no direct constraints on neutral 
current PNC with muons at low energy, we are led to consider two-loop mediation mechanism shown in 
Fig.~\ref{fig:PNC} that transfers parity violation from the muon to the electron sector. Typically, two-loop corrections to PNC amplitudes are not expected to be large. 
However,  because in our model we start with an effective four-fermion
$\bar \psi_p \gamma^\nu \psi_p \bar \psi_\mu \gamma_\nu \psi_\mu$ interaction with a coupling (of mass dimension $-2$) that is much larger than $G_F$ while the 
precision in measuring the weak charge is better than 1\%, one can expect a reasonably strong constraint despite the two-loop 
suppression. 

Currently, the most precise experimental determination of the PNC amplitude for $^{133}$Cs~\cite{Wood:1997zq} is supplemented by 
high-accuracy atomic calculations~\cite{Porsev:2009pr,Dzuba:2012kx} that give a very good agreement of experiment with the SM. For this paper, 
we shall adopt the bound on new physics contribution to the weak charge of cesium nucleus 
at  $2\sigma$ level following the latest theoretical determination~\cite{Dzuba:2012kx},
\begin{equation}
\label{QWlimit}
|{\Delta Q_W}| <0.86.
\end{equation}

Crucially, the $V\gamma\gamma$ vertex generated by the muon loop does not decouple in the limit of $m_\mu\to \infty$, 
because of the properties of the fermionic triangle diagram~\cite{Adler:1969gk}. Moreover, because of what can be viewed as a gauge anomaly, 
there is a sensitivity to the ultraviolet cutoff $\Lambda_{\rm UV}$. Performing the direct calculation of the two-loop induced $V$-electron axial-vector 
coupling in the limit of small momentum transfer and retaining only the logarithmically 
enhanced contributions, we arrive at the following result:
\begin{equation}
\label{Leff}
{\cal L}_{\rm eff} = V_\mu \bar \psi_e \gamma_\nu \gamma_5\psi_e \times \frac{3 \alpha^2 g_A}{2\pi^2} \log\left ( \frac{\Lambda_{\rm UV}^2}{m_\mu^2}\right) .
\end{equation}
Without a UV-complete theory it is then impossible to make a definitive prediction. We note, however, that the simplest way to cutoff the logarithm is to 
extend the model to $\tau$ leptons, and take $g_{A\mu} =- g_{A\tau}$. In that case the answer for the new physics contribution to the 
electron-proton parity violating interaction and the corresponding effective shift of the weak charge of $^{133}$Cs take the following form,
\begin{eqnarray}
{\cal L}_{\rm eff} =  \bar \psi_e \gamma^\nu \gamma_5\psi_e \bar \psi_p \gamma_\nu \psi_p 
\times \frac{ e\kappa}{m_V^2}\times  \frac{3 \alpha^2 g_A}{2\pi^2} \log\left ( \frac{m_\tau^2}{m_\mu^2}\right) ,
\nonumber\\
\Delta Q_W = \frac{12\sqrt{2}\alpha^3}{\pi} \log\left ( \frac{m_\tau^2}{m_\mu^2}\right) 
\frac{\kappa (g_A/e)}{G_Fm_V^2}.~~~~~~~~
\end{eqnarray}

Substituting typical values for the parameters of the model, we arrive at the following shift of the weak charge:
\begin{equation}
|\Delta Q_W| \simeq 1.4\times \left(\frac{Z}{55}\right)\frac{\kappa (|g_A|/e)}{2.5\times 10^{-6}}\left(\frac{10~{\rm MeV}}{m_V}\right)^2.
\end{equation}
While the contribution to $Q_W$ can be either positive or negative, we do not keep track of the sign of $g_A$ since the 
required value for $g_A$ to satisfy $(g-2)_\mu$ can be of either sign, c.f. Eq.~(\ref{cancel}). 

\begin{figure}
\begin{center}
\resizebox{0.95\columnwidth}{!}{\includegraphics{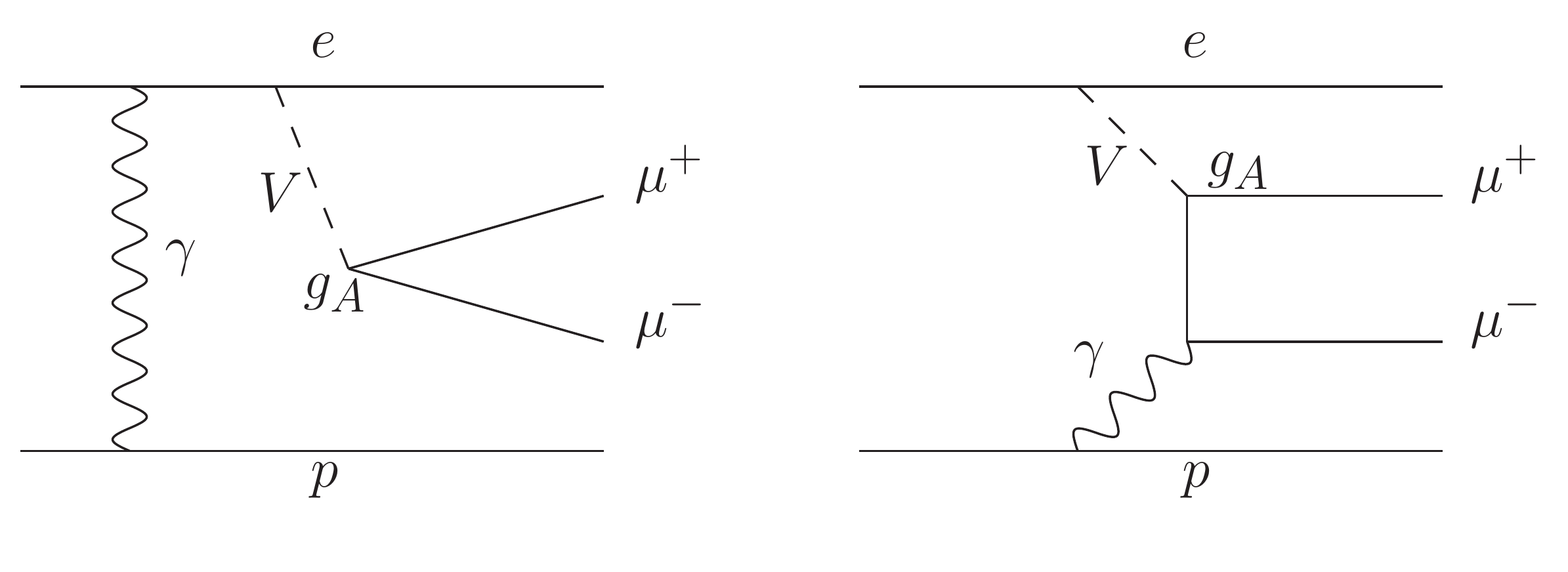}}
\caption{Typical representatives of muon pair production by electron-proton collision 
due to a new force. The parity violation will come about due to the presence of the $g_A$ coupling in the interference with the 
pure QED diagrams.}
\label{fig:muontrident}
\end{center}
\end{figure}

Although atomic parity violation as well as PNC experiments with electron scattering can potentially constrain 
the size of $g_A$, there is also a  question of how to search for enhanced PNC involving muons directly. References~\cite{Batell:2011qq,Missimer:1984hx,McKeen:2012sh} 
have pointed out that polarized muon scattering and muonic atoms can be used for these purposes. 
Here we would like to remark that an alternative way of searching for neutral current PNC with muons is  
polarized electron scattering with  muon pair production, $e_{L(R)} + Z\to e + Z +\mu^+\mu^-$. 
Parity-violating $V$-exchange amplitudes pictured in Fig.~\ref{fig:muontrident} interfere with the QED diagrams,
leading to an asymmetry in the  muon pair-production cross section by the longitudinally polarized electrons,
\begin{equation}
\frac{\sigma_L-\sigma_R}{\sigma_L+\sigma_R} \propto \kappa (g_A/e).
\end{equation}
While the rate for such a process is rather low, new high intensity polarized electron beam facilities
can conceivably be used to search for such an effect.

\section{High-energy constraints}
\label{sec:high-energy}

So far we have dealt with the low-energy observables that are only mildly sensitive to the fact that the 
effective Lagrangian (\ref{start}) at a generic point in $\{g_V,g_A\}$ parameter space does not 
respect the full gauge invariance of the SM. Specifically, we have insisted that the 
muon neutrinos are uncharged under the new force, due to the fact that their interactions are well-known and 
do not have any room for $O(G_F)$ new physics effects, let alone stronger-than-$G_F$ effects as suggested by
the $r_p$ discrepancy. We also do not assume any direct coupling of $V$ to $W$-bosons other than via the 
kinetic mixing $\kappa$. It is then clear that the SM {\em charged} current processes accompanied by the emission of the light
vector boson $V$ from the muon line will be drastically different from a similar process with an emission of a photon. 
In particular, the interaction of the longitudinal part of the $V$ boson will be enhanced with energy
due to the absence of the conservation of the corresponding current. As pointed out in Refs.~\cite{Barger:2011mt,Carlson:2012pc,Carlson:2013mya}, 
direct production of $V$ from muons in $K\to \mu\nu V$ decays can be enhanced by a factor of $m_\mu^2/m_V^2$  
for the $V+A$ current, and even more for the $V-A$ current. In the latter case it is advantageous 
to study very high-energy processes (see e.g. Ref. \cite{Laha:2013xua}), where the enhancement can scale 
as (Energy)$^2/m_V^2$. 

One of the best known charged current processes is the leptonic decay of $W$ boson. 
When $g_V\ne g_A$ (in other words, when the coupling of $V$ boson to the left-handed muon is not zero)
the decay $W\to\mu\nu V$ will be enhanced by $m_W^2/m_V^2$, with the onset of an effectively strong coupling 
when $(g_V-g_A) m_W/m_V \ga 1$. Since this parameter is indeed larger than one for the interesting part of 
parameter space, one should expect a very strong constraint on the model. 
Carrying out explicit calculation in the limit of $g_V\gg g_A$, as implied by $(g-2)_\mu$, and  to leading order in $m_V/m_W$ and $m_\mu/m_W$, we arrive at
\begin{align}\label{Wdec}
\Gamma\left(W\to\mu\nu V\right)&=\frac{g_V^2}{512\sqrt{2}\pi^3}\frac{G_Fm_W^5}{m_V^2} 
\\
&=1.74~{\rm GeV}\left(\frac{g_V}{10^{-2}}\right)^2\left(\frac{10~{\rm MeV}}{m_V}\right)^2.
\nonumber
\end{align}
Because of the prompt decay of $V$ to an electron-positron pair, and the small value of $m_V$, 
this decay will be similar to $W\to \mu \nu \gamma$. 
In any case, the additional channel leads to the increase of the total $W$ width. 
The contribution in Eq.~(\ref{Wdec}) should be compared against the current experimental 
value for the $W$ width, dominated by measurements at the Tevatron~\cite{Wwidth},
\begin{align}
\Gamma_W&=2.085\pm0.042~{\rm GeV}.
\end{align}
Given the agreement of this with SM expectations for $W\to\ell\nu$ and $W\to$hadrons, we limit the contribution of the $\mu\nu V$ mode to the $W$ width to twice its error, leading to a branching
\begin{align}
{\cal B}\left(W\to\mu\nu V\right)<4.0\%
\end{align}
at $2\sigma$. This translates to a limit on the coupling of $V$ to muons of
\begin{align}
g_V<2.2\times10^{-3}\left(\frac{m_V}{10~{\rm MeV}}\right).
\end{align}

\begin{figure}
\begin{center}
\resizebox{0.65\columnwidth}{!}{\includegraphics{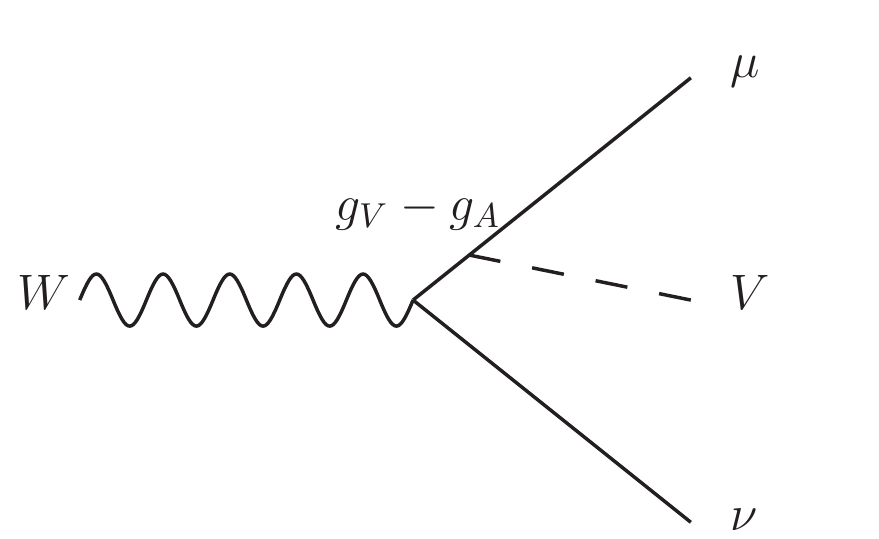}}
\caption{Diagram that leads to the decay $W\to\mu\nu V$.}
\label{fig:3body}
\end{center}
\end{figure}

It is clear that a large correction to $W$ decay is an example of strong high-energy constraints resulting from the lack of the consistent 
SM embedding of the starting point in Eq.~(\ref{start}). There are other processes that can be equally problematic for such models. 
For example, insertion of the virtual $V$ line into the $\mu\nu$ loop in the $W$ self-energy diagram will result in the 
shift of $m_W$ and will impact the very precisely measured $\rho$-parameter of the electroweak theory. Since the lack of the full 
SM gauge invariance, one should expect a power-like sensitivity to the UV cutoff in such theory, which is even stronger enhancement than $m_W^2/m_V^2$. Thus, indeed, these examples show an utmost need for a consistent SM embedding at the level of 
 the very starting point (\ref{start}). 

\section{Combination of all constraints}
\label{sec:combination}

Having performed the required calculations of the  muonium HFS, atomic PNC, and $W$ decays, we are now ready to compile the 
constraints on the parameters of our model. We separate all constraints into low-energy and high-energy ones.

Addressing the low-energy constraints first, it is useful to recall that our model has four parameters, 
$\{m_V,~\kappa,~ g_V, ~ g_A\}$, which enter in the observables in the following combinations, 
\begin{eqnarray}
a_e[m_V,\kappa^2];~~  a_\mu[m_V,(e\kappa+g_V)^2,g_A^2];~~\nonumber\\
\Delta r_p^2[m_V, \kappa g_V];~~\Delta E_{\rm hfs}[m_V, \kappa(e\kappa+g_V)];\\\nonumber
\Delta Q_W [m_V,\kappa g_A];~~ \Delta E_{\rm \mu Mg(Si)}[m_V,\kappa(e\kappa+g_V)]
\end{eqnarray}
The last entry here is the constraint imposed by the agreement of the measured $2p-3d$ transition frequencies in 
muonic magnesium and silicon with the corresponding QED predictions~\cite{Beltrami:1985dc}. 

Besides the indirect constraints on the model via effects induced by virtual $V$, there are, of course, 
direct constraints from the production of $V$ with subsequent decay into $e^+e^-$ pairs, either from $e^+e^-$ colliders or in experiments with fixed targets.
  Thus, searches 
for unexpected spikes in the invariant mass of pairs impose additional constraints on $\kappa$. 
The latest compilations~\cite{Essig:2013lka} show that below $m_V$ of 40 MeV, which is the region of the most interest for us,
$g-2$ of the electron still provides the dominant limits.

In order to present our results in the most concise form, we choose to saturate the constraint coming from $g-2$ of the electron combined 
with atomic determination of $\alpha$. Taking the 2$\sigma$ limit on the maximal deviation of $a_e$ (see, e.g.,~\cite{Davoudiasl:2012ig}), we arrive at maximum allowed 
$\kappa$ for a given value of $m_V$,  Currently, this constraint is given by 
\begin{equation}
\label{max}
|\Delta a_e | \leq 1.64\times 10^{-12} \Longrightarrow \left|\kappa^{\rm max}\right| = 1.8\times 10^{-3}\frac{m_V}{20~{\rm MeV}}.
\end{equation}
The latter equation is valid in the scaling regime $m_V \gg m_e$, but we use the full expression in our numerical treatment. 

\begin{figure}
\begin{center}
\resizebox{0.98\columnwidth}{!}{\includegraphics{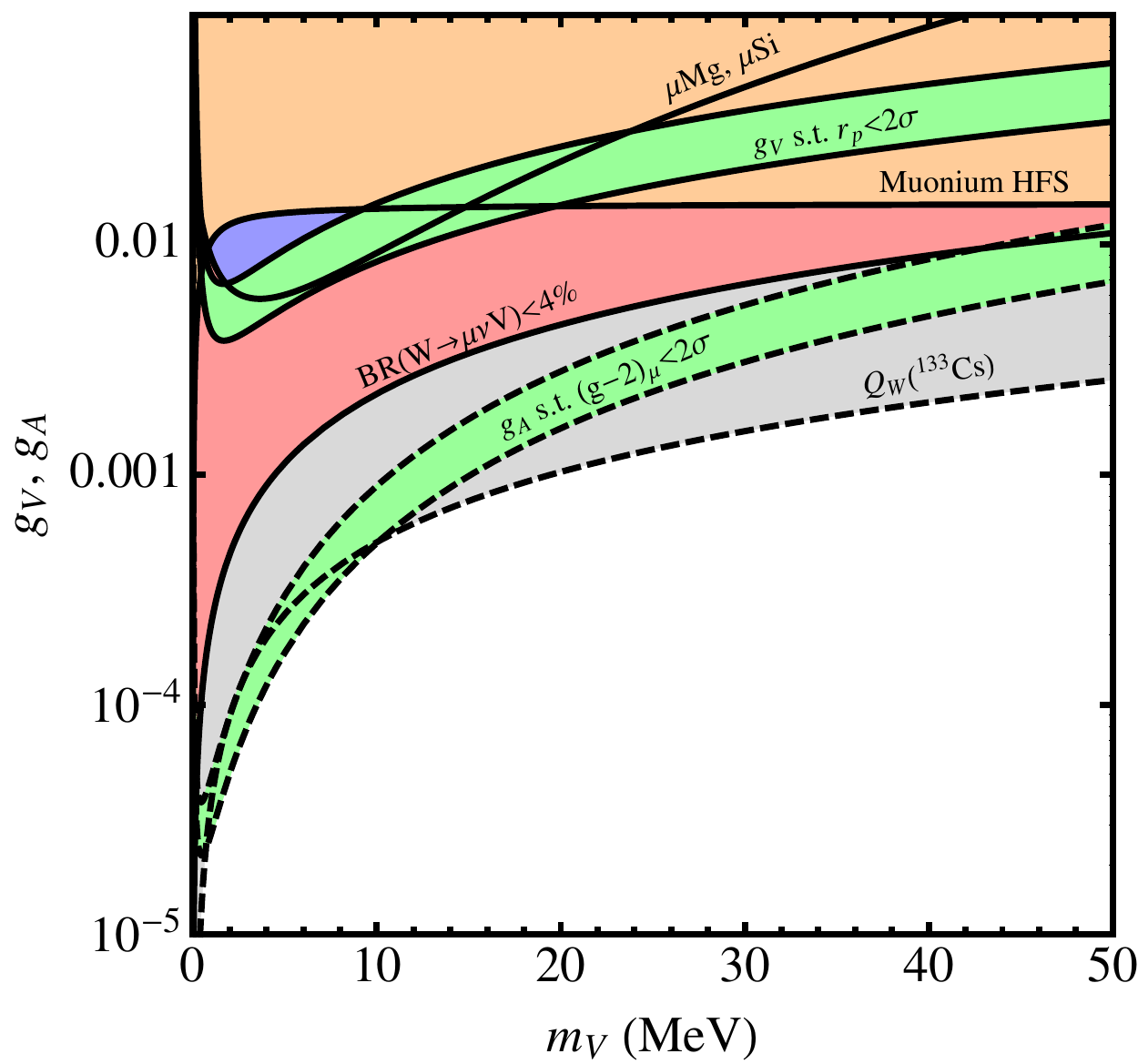}}
\caption{Parameter space of the model, when $\kappa$ is chosen as a function of $m_V$ to saturate the $a_e$ constraint. Solid curves are limits on $g_V$ while dashed ones are limits on $g_A$. The upper green shaded band shows the range of values of $g_V$ that alleviate the $r_p$ discrepancy. The lower green shaded band shows values of $g_A$ required so that $(g-2)_\mu$ theory and experiment agree to $2\sigma$, given values of $g_V$ in the upper band. We show constraints on $g_V$ from muonium HFS, muonic Si and Mg, and $W\to\mu\nu V$ decays (solid curves) and on $g_A$ from PNC in $^{133}$Cs (dashed curve).}
\label{fig:limits} 
\end{center}
\end{figure}

Using this value of $\kappa^{\rm max}$, we determine the required value of $g_V$ that solves the 
$\Delta r_p^2$ discrepancy according to Eq.~(\ref{rp}). Specifically, we require that 
the new physics effect interpreted as  $\left. \Delta r^2\right|_{\mu \rm H} - \left. \Delta r^2\right|_{\rm H}$ is bounded by $2\sigma$ of the CODATA value, 
\begin{equation}
-0.081 ~{\rm fm}^2 \leq \left. \Delta r^2\right|_{\mu \rm H} - \left. \Delta r^2\right|_{\rm H} \leq -0.045 ~{\rm fm}^2.
\end{equation} 
This creates the preferred value for $g_V$, pictured as the upper shaded band with solid borders in Fig.~\ref{fig:limits}. For definiteness we take $\kappa$ to be positive
and for our numerical treatment do not assume $\alpha m_\mu \ll m_V$.

As already stated, such values of $g_V$ are in contradiction with the muon $g-2$ constraints if $g_A=0$. 
Requiring the axial-vector and vector contributions 
to cancel within the $2\sigma$ band around the experimental mean, 
\begin{equation}
1.27\times 10^{-9} \leq \Delta a_\mu(g_V+ e\kappa) + \Delta a_\mu(g_A) \leq 4.47\times 10^{-9},
\end{equation}
we plot the required values of $|g_A|$ as the lower shaded band with dashed borders in Fig.~\ref{fig:limits}. 

As expected, rather small values of the axial-vector couplings, $g_A\ll g_V$, are capable of adjusting the muon $g-2$. However, 
it must be noted that despite the possibility of cancellation, the values of $g_A$ are finely tuned to the values of $g_V$. In other 
words, to every point in the upper band on Fig.~\ref{fig:limits} there is exactly one in the lower band in correspondence. The degree 
of fine-tuning is relatively modest at low values of $m_V$ ({\em e.g.} $\sim 5$\% at $m_V = 3$ MeV) but 
quickly becomes rather extreme as $m_V$ is increased ($\sim$1 part in 1000 at $m_V$ = 30 MeV). 

Besides the $g_V$ and $g_A$ bands, Fig.~\ref{fig:limits} also shows three low-energy exclusion lines: (1) the atomic PNC constraint on $g_A$; 
(2) the muonium HFS constraint on $g_V$; (3) the combination of muonic Mg and Si constraints on $g_V$. The muonium HFS and 
the atomic PNC constraints are given by Eqs. (\ref{hfslimit}) and (\ref{QWlimit}) respectively, while for the muonic Si and Mg we take the weighted mean 
of the  results from Ref.~\cite{Beltrami:1985dc} and allow for a $2\sigma$ deviation,
\begin{equation}
\left| \frac{\Delta E_{3d-2p}}{ E_{3d-2p}} \right| < 6.2 \times 10^{-6}.
\end{equation}

The muonium HFS constraint proves to be 
rather stringent and disfavors all otherwise acceptable values of $m_V$ above $25$ MeV. Notice that while 
$\Delta E_{\rm hfs} $ scales inversely proportional to $m_V$, the constraint line is nearly horizontal because of the 
$\kappa \propto m_V$ choice from Eq.~(\ref{max}).  Muonic Mg and Si are an important constraint, recognized by many 
groups before~\cite{Barger:2010aj,TuckerSmith:2010ra,Batell:2011qq}. It reduces the allowed parameter 
quite significantly, but is unable to close it.  (Around $m_V\sim 10$ MeV, for example, 
the model can still be consistent with all the constraints at $\sim 2\sigma$.) The atomic PNC constraint is also very 
sensitive to $g_A$, 
disfavoring all solutions with $m_V>10$ MeV. We conclude, though, that the combination of all low-energy constraints cannot 
decisively exclude the new muonic vector force solution to the $r_p$ puzzle. 

However, the addition of the high-energy constraints changes the story: Fig. \ref{fig:limits} shows that 
the $W\to \mu\nu V$ constraint is a factor of a few below the $g_V$ band over the entire parameter range of interest. 
This is a direct consequence of choosing a zero coupling of $V$ to neutrinos, $W$ bosons and $g_V \neq g_A$.

\section{Discussion and conclusions}
\label{sec:conclusions}

By combining all the constraints, we can assess the phenomenological status of the model
with a new ``dark photon" type vector force with additional couplings to the muon. Our main conclusion is that 
the model designed to ``remove" $r_p$ anomaly by adjusting $g_V$ and $g_A$ couplings
survives current generation of the low-energy constraints,  but fails the high-energy tests because of the 
lack of a consistent SM embedding. 
The $g-2$ of the muon also requires some fine tuning. 
The tuning is minimized in the mass region of very light, $m_V<10$ MeV, mediators,
and it is logical to conclude that this is the preferred mass range for the model. 

We provide some further comments below. 

\begin{itemize}

\item The biggest challenge for models of the kind considered here is their embedding into the 
SM. In particular, $g_V \gg g_A$ case can be interpreted as mostly $R+L$  coupling. The presence of 
significant left-handed component implies large couplings of neutrinos to a new force, which is 
incompatible with the strength of the force considered here. Therefore, these types of models, unfortunately, 
remain rather artificial: because of their inability to deal with neutrinos, they are subject to strong
high-energy constraints due to (Energy)$^2/m_V^2$ enhancement. The gauged right-handed muon model of 
Ref.~\cite{Batell:2011qq} has the least number of pathologies and is not constrained by high-energy processes
as $g_V=g_A$, but it requires additional contributions to allow for stringent tunings of the muon $g-2$ and the atomic PNC. 

\item The fact that muonic HFS turns out to be rather constraining is encouraging, giving the 
fact that a new generation of experiments is being planned. 

\item  The results of the muonic deuterium Lamb shift measurements are about to be released soon by the same group that  
measured $\mu$H. Together with the isotopic shift constraints and accurate theoretical calculations of deuterium polarization, 
this measurement will be able to shed some additional light on the internal self-consistency of the results. In the speculative world 
of new physics models, it will provide extra invaluable information on whether an additional coupling of new forces to neutrons
is warranted. In particular, it will clarify whether one should revisit constraints on new scalar-mediated forces. 

\item It is worth emphasizing that the new muon-proton scattering experiment at PSI, MUSE~\cite{Pohl:2013yb}, may not detect the 
presence of the new muon-specific force if the mediator mass is small. Indeed, the experiment will use a momentum transfer of 
$O(100)$ MeV, which is larger than the preferred mediator mass range. Consequently, the measured charge radius 
of the proton in the muon-proton scattering may not differ from $e-p$ result {\em despite} the possible presence of a new force. 

\item Direct production of new particles with their subsequent decay to electron-positron pairs may be efficiently searched with the 
muons in the initial state. For example, a careful study of $K\to \mu\nu e^+e^-$ may reveal unexpected peaks at low invariant mass of 
the pair. A new experiment searching for $\mu\to 3e$ decay 
\cite{Blondel:2013ia} will also have capabilities of probing $\mu\to e\bar \nu\nu V$ decays. 

\end{itemize} 

We believe that future progress in gaining understanding of the $r_p$ problem will come from experiment. 
Besides the previously mentioned results with the muonic deuterium and the ongoing experiment with muonic helium, 
one should pay close attention to improvements in experiments with spectroscopy of ordinary hydrogen. 
If subsequent muonic experiments show no particular anomalies, while the new results with ordinary hydrogen reinforce 
the $r_p$ problem, it may be worth checking the idea of the electron-specific force, that creates a small amount of {\em repulsion}
between electron and a proton, and have a mass of the mediator in the range between $\alpha m_e$ and $m_e$. 
In this case, one can avoid the constraints from $g-2$ of the electron, as the required scale of couplings will be tiny.
But at the same time, the question of consistent embedding of such new force into the SM will still remain, and such models 
will face the very same difficulties as ``muonic forces" discussed in this work. In addition, sub-MeV mediator masses
are also subject to very strong constraints from cosmology and astrophysics. 

\begin{acknowledgements}
MP would like to thank A.~Antognini, B.~Dasgupta, 
R.~Pohl, M.~Strassler, and I.~Yavin for useful discussions. 
SGK is supported by DFG under grant HA~1457/9-1.
DM is supported by DOE Grant No. DE-FG02-96ER40956 and acknowledges past support during early stages of this work from NSERC.
Research at the Perimeter Institute is supported in part by the
Government of Canada through NSERC and by the Province of Ontario through MEDT.
\end{acknowledgements}

\section{Appendix: full answer for muonium hfs}

The contribution of an additional vector force to muonium hfs can be extracted from the 
previous calculations that employed a massive vector in the 
evaluation due to hadronic vacuum polarization contribution~\cite{plb}. 
The contribution of a massive vector reads
\begin{equation}
\frac{\Delta E_{\rm hfs}}{E_{\rm hfs}} = 2 \frac{\alpha^\prime}{\pi} \frac{m_e}{m_\mu}K_{\rm Mu}(s),
\end{equation}
where $K_{\rm Mu}(s)$ is discussed below in detail. Here
\[
s=m_V^2
\]
and $\alpha^\prime$ is the coupling for the new vector force.

In the limit
\[
\frac{m_V}{2m_\mu}\ll1
\]
one finds
\begin{equation}
K_{\rm Mu}(s\to 0) \to \frac{4\pi m_\mu}{m_V}\;,
\end{equation}
leading to the result of Eq.~(\ref{resulthfs}). 

The kernel $K_{\rm Mu}(s)$ was investigated in~\cite{plb}. We are interested in $m_V$ around $10\;$MeV 
and thus corrections to the leading term of order $m_e/m_V$ and $m_V/m_\mu$ should be under control.
The full expression for the kernel is~\cite{plb}
\begin{align}
K_{\rm Mu}(s) &= -\left(\frac{s}{4m_\mu^2}+2 \right)
\sqrt{1-\frac{4m_\mu^2}{s}}\log\frac{1+\sqrt{1-\frac{4m_\mu^2}{s}}}{1-\sqrt{1-\frac{4m_\mu^2}{s}}}
\nonumber
\\
&\quad\quad+\left( \frac{s}{4m_\mu^2} +\frac{3}{2} \right)
\ln{\frac{s}{m_\mu^2}} -\frac{1}{2} \;.
\label{plb:1}
\end{align}
Analytically continuing to $s<4m_\mu^2$, this becomes
\begin{align}
K_{\rm Mu}(s) &= 2\left(\frac{s}{4m_\mu^2}+2 \right)
\sqrt{\frac{4m_\mu^2}{s}-1} \,\tan^{-1}\sqrt{\frac{4m_\mu^2}{s}-1}
\nonumber
\\
&\quad\quad+\left( \frac{s}{4m_\mu^2} +\frac{3}{2} \right)
\ln{\frac{s}{m_\mu^2}} -\frac{1}{2} \;.
\end{align}

\begin{figure}
\begin{center}
\resizebox{0.9\columnwidth}{!}{\includegraphics{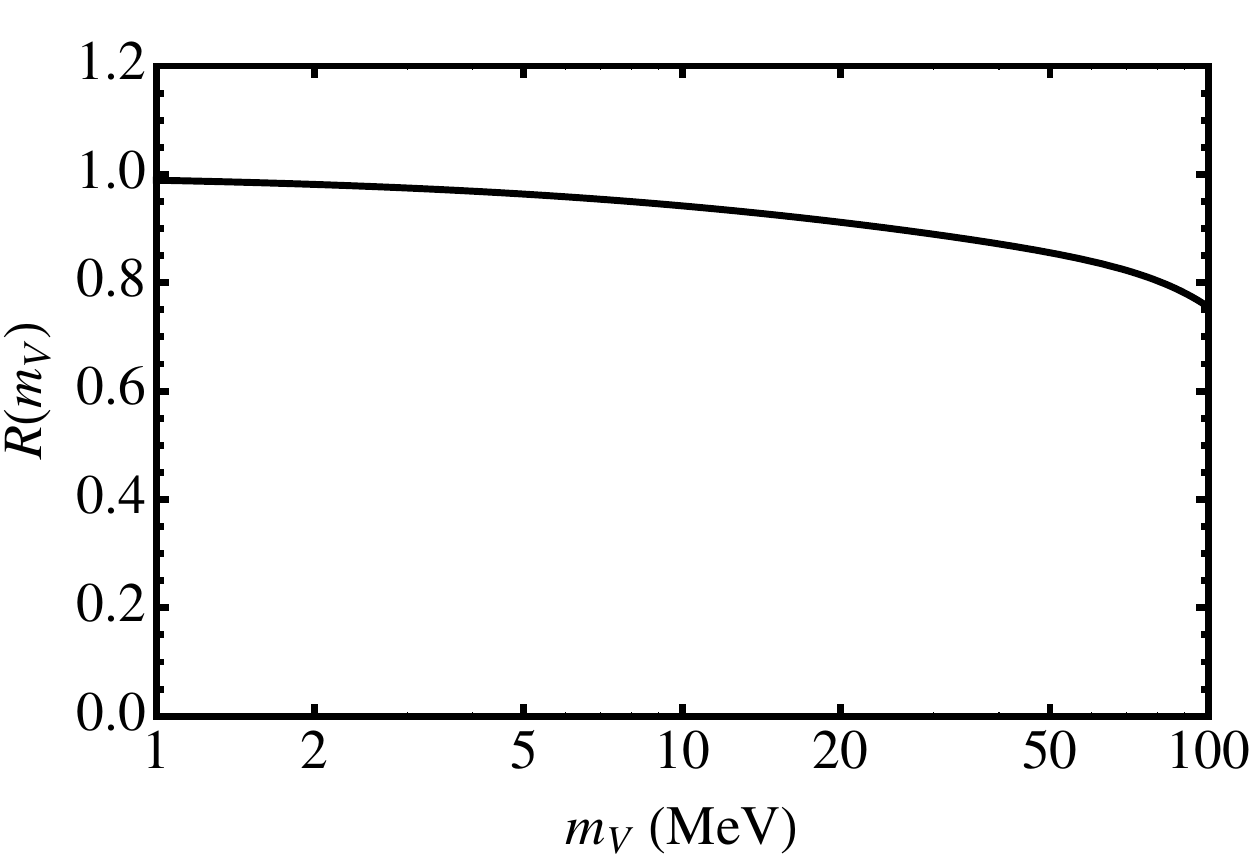}}
\caption{$R(m_V)$ as defined in Eq.~(\ref{eq:RmV}) over the relevant range of $m_V$.}
\label{f:R}
\end{center}
\end{figure}

Normalized on its value at $m_V=0$, we define the correction factor $R(m_V)$,
\begin{align}
R(m_V)&=\frac{K_{\rm Mu}(s=m_V^2)}{K_{\rm Mu}(0)}
\label{eq:RmV}
\\
&= \frac{m_V}{4\pi m_\mu} \times K_{\rm Mu}(s=m_V^2).
\nonumber
\end{align}
Eq.~(\ref{resulthfs}) is then generalized to
\begin{equation}
\frac{\Delta E_{\rm hfs}}{E_{\rm hfs}} = \frac{8 \alpha' m_e}{m_V}R(m_V),
\end{equation}
at arbitrary values of $m_V/m_\mu$. Over the entire mass range of interest, $R(m_V)$ varies from unity by less than about 20\%, as shown in Fig.~\ref{f:R}.

\end{document}